\documentstyle[aps,eqsecnum,epsfig]{revtex}
\begin{document}
\draft
\preprint{}
\title{Phase Transition and Hybrid Star in a Nonlinear $\sigma-\omega$ Model}
\author{S. Acharya$^\dagger$, L. Maharana$^{\dagger,*}$,
R. Mohanty$^\dagger$ and P. K. Panda$^\ddagger$}
\address{$^\dagger$Physics Department, Utkal University, 
Bhubaneswar-751004,India.\\
$^\ddagger$Non-Acceleration Particle Physics Group, Indian Institute of 
Astrophysics, Bangalore-560034,India\\
$^*$FB Physik, Universit\"at Kaiserslautern, D-67653 Kaiserslautern, 
Germany}
\maketitle
\begin{abstract}
The phase transition between the nuclear matter and
the quark matter is examined. The relativistic mean field theory(RMF) is
consider with interacting nucleons and mesons using TM1 parameter set for the
nuclear matter equations of state. It is found that the trasition point 
depends on coupling constant $\alpha_s$  and bag pressure. 
From the study of the structure of a hybrid neutron star, it is
observed that the star contains
quark matter in the interior and neutron matter on the outer perifery.
\end{abstract}

\section {Introduction}
It is widely believed that nuclear matter undergoes a phase 
transition to quark matter at high densities and/or high 
temperatures. The high temperature limit is expected
to have interesting consequences in heavy ion collision and/or in
cosmology, whereas high baryon density behaviour is important for
the study of neutron stars.

It is expected that Quantum Chromodynamics (QCD) as the fundamental theory
of the strong interaction  should  explain
possible  modifications of hadron properties in the nuclear
medium. However, typical nuclear phenomena at intermediate and 
low energies cannot be analytically derived from QCD although
one hopes that QCD will be solved numerically on the lattice in
near future. Meanwhile we are left with the construction of
phenomenological models in order to try to describe nuclear 
phenomena and its bulk properties. Walecka and others \cite{Walecka,Serot}
used a kind of relativistic scalar-vector theory to describe
the nucleon-nucleon properties of nuclear matter as well as of 
properties of finite nuclei.
Some of the drawbacks of the original Walecka model are that the 
effective nucleon mass obtained at high densities is too
small and its incompressibility at the energy density saturation
is too large. To eliminate these difficulties Boguta and Bodmer \cite{Boguta}
modified the original model by introducing selfcoupling terms to the 
scalar field. The inclusion of nonlinear terms to the scalar field
surprisingly improved the results of nuclear matter as well as of
finite nuclei. However, in most of the successful parameter
sets, the last term of the self coupling constant is found to be
negative \cite{Rufa}. This negative value of the last term gives an unphysical
situation at high density, which is essential for a further modification
of the model. This modification is done by Bodmer \cite{Bodmer}
by introducing a quartic nonlinear term to the vector potential
to study the equation of state. Later on this suggestion was 
considered to study the properties of finite nuclei \cite{Toki},
which gives the nuclear matter and finite nuclei properties excellently well
\cite{Gmuca}.

Many of the existing parametrisation
is unable to reproduce the properties of finite nuclei, nuclear matter
and that of the physical properties of accreting steller objects,
like neutron star etc. Recently a detail calculation has been done with
different models to study the properties of neutron stars \cite{Prakash}.
 Glendenning \cite{Glen} has studied the properties
of neutron star in the framework of nuclear relativistic field theory.
This parameter set is not applicable directly to finite nuclei. Similarly
many parameter sets which explain the properties of accreting matter is
unable to explain the properties of finite nuclei as well as of normal
nuclear matter properties. On the other hand, the parameter set which is
able to explain the properties of finite nuclei and normal nuclear
matter properties, failed to explain the properties of accreting
matter. In this work, our aim is to see the applicability of the
improved parametrisation of Sugahara and Toki.\cite{Toki} to nuclear matter 
which explains well the properties of finite, including superheavy 
nuclei, infinite nuclear matter and the properties of negative energy
bound states at normal as well as at high densities \cite{pkp}.

In a similar study~\cite{mmpp}, the equation of state for neutron matter
is obtained in a nonperturbative method with pion dressing of neutron matter,
an analysis similar to that of symmetric nuclear matter. 
The quark matter sector was treated perturbatively with bag constant 
$B^{\frac{1}{4}}$ = 148 GeV. Stable solution for such a quark-neutron 
hybrid star was obtained with Chandrasekhar limit as 1.58$M_{\odot}$ and 
radius around 10 km. However in the present calculation we have 
considered the effect of $\sigma - \omega$ mesons with the nonlinear
interactions and have observed
the increase in size and mass of such stars with mass about 2.3$M_{\odot}$
and radius 13.5 km.

In a further study, Sugahara and Toki~\cite{sut} have taken $\Lambda - \omega$
tensor coupling and found a heavier critical mass of neutron star beyond
observational boarder but without tensor coupling they have shown that their
result does not agree with observational result indicating that the tensor coupling
is indespensable for meeting observational requirement.

We consider here relativistic mean field theory with interacting 
nucleons and mesons, using a nonlinear version in both $\sigma$
and $\omega$ mesons for the nuclear matter equation of states.
Quark matter is treated perturbatively for high densities at short
distances \cite{ellis}. A first order phase transition between
nuclear matter and quark matter seems to be indicated. Solutions
of the Tolman - Oppenheimer - Volkoff (TOV) equations yield a
hybrid star having a quark core with a crust of neutron matter.

The paper is organized as follows. In Sec. II, we present a brief
theory for nuclear matter (neutron matter) equation of state.
The quark matter equation of state is
discussed in Sec. III. In section IV, we discuss the structure of
hybrid neutron star. A summary and concluding remarks are
given in Sec. V. 

\section {Nuclear matter equation of state}
We start with the effective 
Lagrangian density for a nucleon-meson many-body system
for nuclear matter. In this Lagrangian we have considerd only
the interction of nucleons with $\sigma$ , $\omega$ and $\rho$ mesons. 

The  Lagrangian is given
as \cite{Walecka,Rufa,Toki}
\begin{eqnarray}
{\cal L}&=&\bar\psi(i\gamma^\mu\partial_\mu-M)\psi
+{1\over 2}\partial^\mu\sigma\partial_\mu\sigma-
{1\over 2}m_\sigma^2\sigma^2+{1\over 3}g_2\sigma^3
+{1\over 4}g_3\sigma^4 -g_s\bar\psi\psi\sigma\nonumber\\
&-&{1\over 4}\Omega^{\mu\nu}\Omega_{\mu\nu}
+{1\over 2}m_\omega^2 \omega^\mu \omega_\mu
+{1\over 4}c_3(\omega_\mu\omega^\mu)^2-g_\omega\bar\psi
\gamma^\mu\psi \omega_\mu-{1\over 4}R_{\mu\nu}^a R^{a\mu\nu}\nonumber\\
&+&{1\over 2}m_\rho^2 R_\mu^a R^{a\mu}-g_\rho\bar\psi\gamma^\mu
\tau^a\psi R^{a\mu}
\end{eqnarray}
The fields for the $\sigma$,$\omega$ and $\rho$-mesons are denoted by $\sigma$, 
 $\omega_{\mu}$ and $R\mu$ respectively and $\psi$ is the Dirac spinor  
for the nucleon. Here $g_{s}$, $g_{w}$, $g_{\rho}$ are the coupling
constants for $\sigma$, $\omega$ and $\rho$- mesons
and $g_2$, $g_3$ and $c_3$ are self coupling
constants. M is the mass of the nucleon  and
$m_{\sigma}$, $m_{\omega}$ and $m_{\rho}$  masses of the $\sigma$, 
$\omega$ and ${\rho}$-mesons respectively.The contribution of $\rho$-mesons
to neutron matter is essential and has effect on the formation of 
hybrid stars .

In Ref.\cite{Gmuca} it has been shown that RMF approach is successful 
to describe the result of Relativistic Dirac Bruckner Hartree-Fock (RDBHF) 
calculations in nuclear matter. It  is found that although the RMF model with 
scalar self-interactions
is able to describe effectively the binding energy of nuclear matter as well
as the bulk properties of finite nuclei, this is not followed by a 
proper description of the effective nucleon mass and a time-like 
component of the vector self-energy. This is caused mainly by a 
too restrictive treatment of $\omega-$meson term in the RMF approach,
which does not take into account the density dependence produced
by the relativistic Dirac-Brueckner approximations to all
mesons involved in the theory. The study of finite nuclei and
nuclear matter of  
Sugahara and Toki \cite{Toki} shows that the vector potential of the RMF theory
increases linearly with density and gets stronger, while  
RDBHF bends down with
density. The scalar potential  of the RMF theory seems to be
overestimating the RDBHF  results at high density in order to
compensate for the strong repulsion in the vector channel. This
is the reason for providing the wrong sign in $\sigma^4$ self-coupling
constant in most of the successful parameter sets. Thus, Sugahara
and Toki introduced a nonlinear term ($\omega_\mu \omega^\mu)^2$
into the $\omega$ vector meson potential to study the
properties of finite nuclei.

In the mean field approximation the meson field operators are replaced by 
their expectation values. We also consider the isotropic system at rest.
The equation of motions for meson and nucleon fields are 
\begin{mathletters}
\begin{eqnarray}  
m^2_\sigma\sigma =-g_s\rho_s-g_2\sigma^2-g_3 \sigma^3 \label{gs}\\
m^2_\omega\omega_0 = g_\omega\rho_B-c_3 \omega_0^3\\
m^2_\rho R_{0}^3 = { 1\over 2}g_\rho<\bar\psi\gamma^0\tau_3\psi>\\
m^{2}_\rho R_{03} = g_{\rho} \rho_{03}
\end{eqnarray}
where
\begin{eqnarray}
\rho_{03}={1\over 2}<\bar\psi\gamma^0\tau_3\psi>
\end{eqnarray}
and
\begin{eqnarray}
(-i~\vec\alpha\cdot\vec\bigtriangledown
&+&\beta M^*)\psi = (E - g_\omega \omega_0)\psi
\end{eqnarray}
\end{mathletters}
In the above  we use the effective nucleon mass $M^*=M+g_\sigma\sigma$. 
The source terms
for scalar and vector fields are the scalar density $\rho_s=<\bar \psi \psi>$
and the vector(baryon) density $\rho_B=<\psi^\dagger\psi>$ respectively.
Using the standard positive energy solutions of the Dirac equation,
we obtain
\begin{equation}
\rho_s =\frac{\gamma}{2\pi^2}\int_0^{k_f} k^2
\frac{M^*}{\sqrt{k^2+{M^*}^2}} dk
\label{rhos}
\end{equation}
and 
\begin{equation}
\rho_B=\frac{\gamma}{6\pi^2}k_f^3
\end{equation}
Here we assume that nuclear matter consists of filling nucleon levels up to
the Fermi momentum $k_f$ and $\gamma$ is the spin-isospin 
degeneracy factor ($\gamma=4$ for
nuclear matter and $\gamma=2$ for neutron matter). The effective nucleon mass
$M^*$ has to be determined self-consistently at each density by solving
equation (\ref{gs}) for the scalar field and (\ref{rhos}) 
for the scalar density.

The energy density of the nuclear matter in the mean field approach is
given by
\begin{equation}
\epsilon=\epsilon_N+\epsilon_\sigma+\epsilon_\omega+\epsilon_\rho
\label{enucl}
\end{equation}
Here $\epsilon_N$ is the energy density of nucleons of mass $M^*$
\begin{mathletters}
\begin{eqnarray}
\epsilon_N &=&\frac{\gamma}{2\pi^2}\int_0^{k_f}k^2{\sqrt{k^2+{M^*}^2}}
dk\nonumber\\
&=&\frac{\gamma}{2\pi^2}\left[\frac{1}{4}k_f(k_f^2+{M^*}^2)^{3/2}
-\frac{1}{8}{M^*}^2 k_f{\sqrt{k_f^2+{M^*}^2}}
-\frac{1}{8}{M^*}^4 ln \left( \frac{k_f+{\sqrt{k_f^2+{M^*}^2}}}{M^*}
\right)\right]
\end{eqnarray}
The energy density $\epsilon_\sigma$ is the sigma meson interaction term
which may be written as
\begin{equation}
\epsilon_\sigma=\frac{1}{2}m_\sigma^2\sigma^2+\frac{1}{3}g_2\sigma^3
+\frac{1}{4}g_3\sigma^4
\end{equation}
The third term $\epsilon_\omega$ is the omega-meson interaction term 
which is given by
\begin{equation}
\epsilon_\omega =-\frac{1}{2}m_\omega^2\omega_0^2-\frac{1}{4}c_3\omega_0^4
+g_\omega \omega_0 \rho_B
\end{equation}
and
\begin{equation}
\epsilon_\rho =\frac{1}{2}m_\rho^2 R_{03}^2
\end{equation}
\end{mathletters}

Similarly the pressure for the nuclear matter is given by
\begin{equation}
P=P_N+P_\sigma+P_\omega+P_\rho
\label{pnucl}
\end{equation}

\begin{mathletters}
\begin{eqnarray}
P_N &=&\frac{1}{3}\frac{\gamma}{2\pi^2}\int_0^{k_f}
\frac{k^4}{\sqrt{k^2+{M^*}^2}} dk\nonumber\\
&=&\frac{1}{3}\frac{\gamma}{2\pi^2}\left[\frac{1}{4}k_f^3{\sqrt{k_f^2+{M^*}^2
}}-\frac{3}{8}{M^*}^2 k_f{\sqrt{k_f^2+{M^*}^2}}
+\frac{3}{8}{M^*}^4 ln \left( \frac{k_f+{\sqrt{k_f^2+{M^*}^2}}}{M^*}
\right)\right]
\end{eqnarray}

\begin{equation}
P_\sigma=-\frac{1}{2}m_\sigma^2\sigma^2-\frac{1}{3}g_2\sigma^3
-\frac{1}{4}g_3\sigma^4
\end{equation}

\begin{eqnarray}
P_\omega &=& -\epsilon_\omega+\rho_B\frac{\partial\epsilon_\omega}{\partial
\rho_B}\nonumber\\
&=&\frac{1}{2}m_\omega^2\omega_0^2+\frac{1}{4}c_3\omega_0^4+g_\omega\rho_B^2
\left[\frac{g_\omega}{m_\omega^2+3c_3 \omega_0^2} \right]
-\rho_B(m_\omega^2\omega_0+c_3\omega_0^3)
\left[\frac{g_\omega}{m_\omega^2+3c_3 \omega_0^2} \right]
\end{eqnarray}

\begin{equation}
P_\rho=\frac{1}{2}m_\rho^2 R_{03}
\end{equation}
\end{mathletters}
Here we use TM1 parameter set. The values of the parameter set are 
\cite{Toki} M = 938.0, $m_\sigma$ = 511.198, $m_\omega$ = 783.0,$m_\rho$ = 770,
$g_s$ = 10.0289, $g_\omega$ = 12.6139,$g_\rho$ = 4.6322 $g_2$ = $-7.2325$ fm$^
{-1}$,$g_3$ = 0.6183 and $c_3$ = 71.3075 where masses are in MeV. The 
corresponding nuclear matter properties obtained from the parameter set are
$\rho_0$ = 0.145 fm$^{-3}$, $E/A = -16.3$ MeV, K = 281 MeV, $M^*/M = 0.634$
and $a_{asy}$ = 36.9 MeV, where $\rho_0$, $a_{asy}$, K, and E/A
are the density, asymmetric parameter, compressibility modulus
and the binding energy per particle, respectively. Sugahara and Toki
\cite{Toki} show that the TM1 parameter set gives more closer
results with the RDBHF than the NL1 and NL-SH parameter sets. Also,
we know earlier \cite{Serot} that the linear set of Horowitz and Serot
gives stiff equation of states and predicts a too
high value of compressibility modulus of about 560 MeV whereas the 
empirical value is  $210\pm 30$ MeV \cite{Blaizot}. The value obtained
by TM1 parameter set is more convincing. In figure 1, we have shown the 
behaviour of pressure against density in TM1 and Nl-SH parameter set.

\begin{figure}
\centerline{\psfig{figure=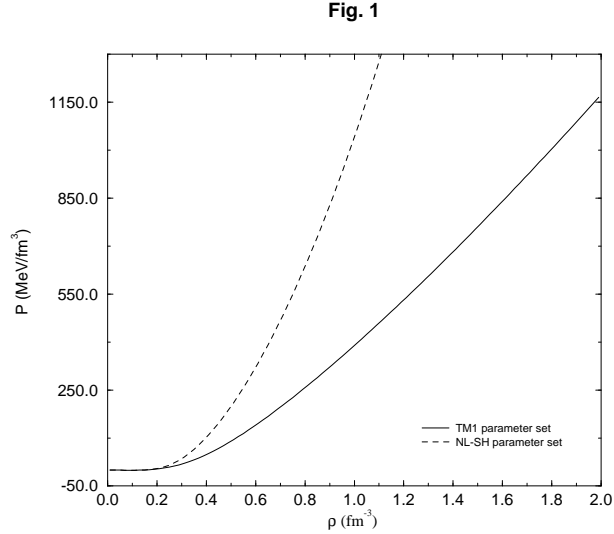,width=3in,angle=-90}}
\caption{Pressure (P) versus density ($\rho$) of nuclear
matter is shown with TM1 parameter set. }
\end{figure}

\section{Quark matter equation of state and phase transition}
Existence of quark matter in the core of neutron stars/pulsars is
an exciting possibility \cite{Witten}.
Densities of these stars
are expected to be high enough to force the hadron constituents or
nucleons to overlap thereby yielding quark matter. Since the
distance involved is small, perturbative QCD is used to derive
quark matter equation of state. We take the quark matter equation
of state as in Refs. \cite{kapusta,freedman} in which u,d and s 
quark degrees of freedom are included in addition to electrons.
Here we set the electron, up and down 
quark masses to zero \cite{kapusta} and
the strange quark mass is taken to be 180 MeV. In chemical 
equilibrium $\mu_d=\mu_s=\mu_u+\mu_e$. In terms of baryon and
electric charge chemical potentials $\mu_B$ and $\mu_E$, one has
\begin{equation}
\mu_u={1\over 3}\mu_B+{2\over 3}\mu_E,\quad
\mu_d={1\over 3}\mu_B-{1\over 3}\mu_E,\quad
\mu_s={1\over 3}\mu_B-{1\over 3}\mu_E.
\end{equation}

The  pressure contributed by the quarks is computed to order
$\alpha_s=g^2/4\pi$ where $g$ is the QCD coupling constant.
Confinement is simulated by a bag constant B. The electron pressure is
\cite{kapusta}
\begin{equation}
P_e={\mu_e^4\over 12\pi^2}.
\end{equation}
The pressure for quark flavor f, with f=u,d or s is 
\cite{ellis,kapusta,freedman}
\begin{eqnarray}
P_f&=&{1\over 4\pi^2}\left[\mu_f k_f(\mu_f^2-2.5m_f^2)+
1.5m_f^4 ln\left({\mu_f+k_f\over m_f}\right)\right]\nonumber\\
&-&{\alpha_s\over \pi^3}\left[ {3\over 2}\left(\mu_f k_f-m_f^2
ln \left({\mu_f+k_f\over m_f}\right)\right)^2-k_f^4\right].
\end{eqnarray}
The Fermi momentum is $k_f=(\mu_f^2-m_f^2)^{1/2}$. The total pressure,
including the bag constant B is
\begin{equation}
P=P_e+\sum_f P_f-B.
\label{pquark}
\end{equation}
There are only two independent chemical potentials $\mu_B$ and
$\mu_E$. $\mu_E$ is adjusted so that the matter is electrically
neutral, i.e. $\partial P/\partial \mu_E=0$. The baryon number density
is given by $\rho=\partial P/\partial \mu_B$.

We now consider the scenario of phase transition from nuclear
matter to quark matter. As usual, the phase boundary of the 
coexistence region between the nuclear and quark phase is determined
by the Gibbs criteria. The critical pressure and critical chemical potential
are determined by the condition
\begin{equation}
P_{nm}(\mu_B)=P_{qm}(\mu_B).
\end{equation}
We take $\alpha_s=0.5,\; 0.6$ and the bag constant $B=(150\;
\mbox{MeV})^4,\; (155\;\mbox{MeV})^4$,
which is a reasonable value to calculate pressure in the quark sector.
Schaab et al. \cite{Schaab} used this value to be $B^{1/4}=145$ MeV. 
However in a calculations of Glendinning \cite{Glen1}  the bag pressure
was taken as $B^{1/4}=180$ MeV. In this calculations\cite{Glen1} the  transition
was determined for the above bag constant which places the energy per baryon
of strange quark matter 1100 MeV, well above the energy per nucleon in infinite
nuclear matter as well as the most stable nucleus, $^{56}$Fe 
(E/A$\approx 930$ MeV).In figure 2,we have plotted pressure versus
chemical potential for nuclear matter and quark matter.The solid line is shown 
for nuclear matter,with TM1 parameter set. The dash and the dotted lines
are shown for quark matter with
$\alpha_s=0.5$ and $\alpha_s=0.6$ with bag pressure $B=(155\; \mbox{MeV})^4$  
respectively . A remarkable feature of the state of affair is that there exist
transition points for nuclear matter to quark at diffrent pressures and chemical 
potentials . These trasition points
($P_{crit}, \mu_{crit}$) are at $\alpha_s=0.6$ with (150 MeV/fm$^3$, 1280 MeV)   
and at $\alpha_s=0.5$ with (260 MeV/fm$^3$, 1445 MeV)
showing dependance on $\alpha_s$ and these also indicate the first order
phase transition from nuclear matter to quark matter at differnt thermodynamical  
 conditions . We also note that the phase transition seems to occur
around the number density of about 5 times the nuclear matter
density.These points also change under differnt bag pressure. Figure 3 shows the   
phase diagram with a different bag pressure $B=(150\; \mbox{MeV})^4$. Here we 
found that transition point decreases with decrease of bag pressure, whereas 
the transition point shifts to a higher value with an increase in coupling
constant $\alpha_s$.The early phase transition from nuclear matter
to quark matter obviously implies that the interior of ``neutron star" will
usually consists of quark matter. We investigate this possibility
in the next section.

\begin{minipage}[t]{2.7in}
\begin{figure}
\psfig{figure=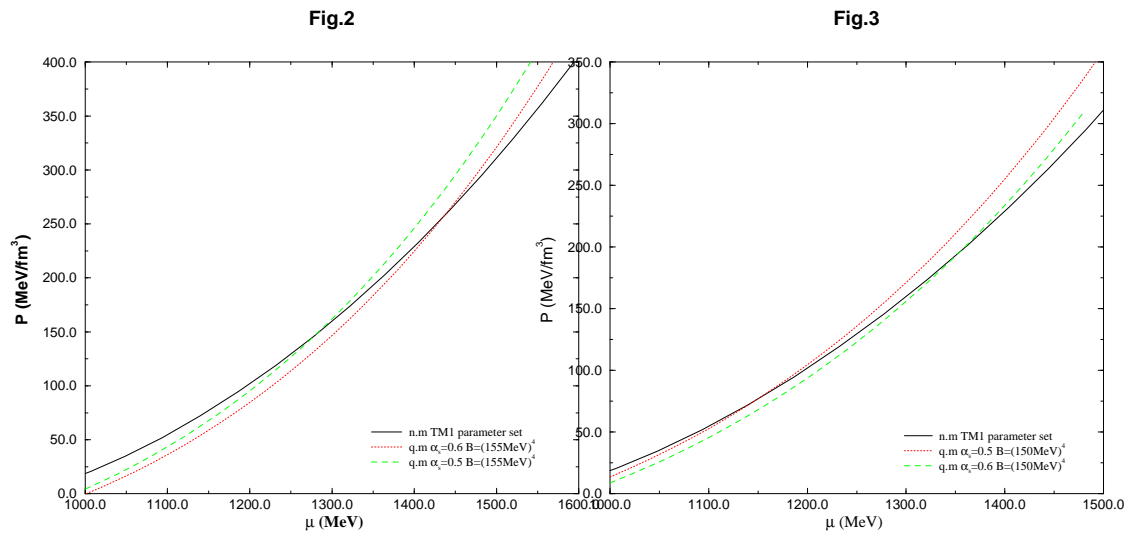,width=3in,angle=-90}
\caption{ Pressure (P) versus chemical potential ($\mu$) for nuclear
matter and for quark matter with various $\alpha_s$ at 
constant bag pressure (B).}
\end{figure}
\end{minipage}
\begin{minipage}[t]{2.7in}
\begin{figure}
\psfig{figure=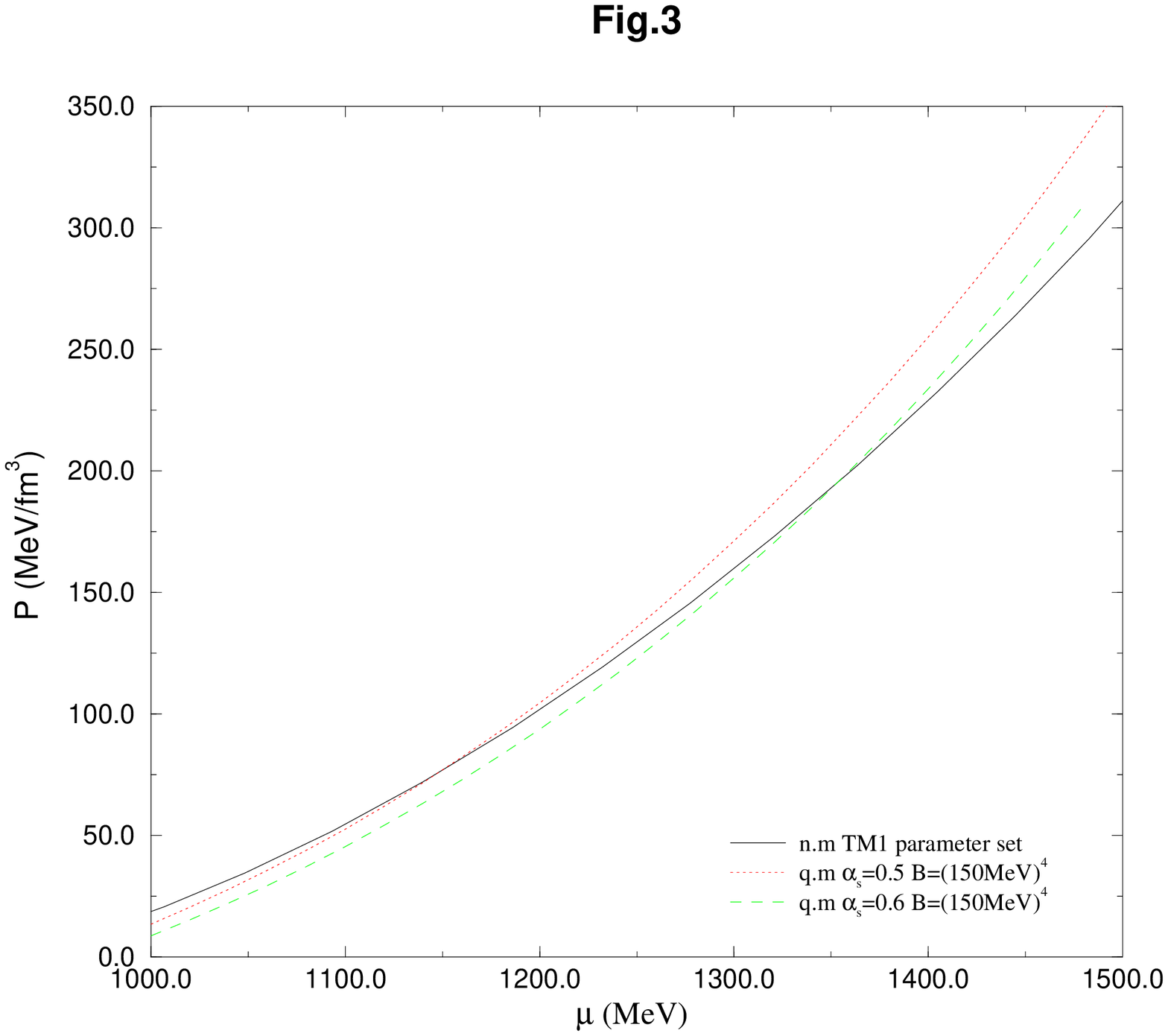,width=3in,angle=-90}
\caption{ Same as Fig.2 with a different bag pressure (B).} 
\end{figure}
\end{minipage}

\section{Hybrid stars}
For the description of neutron star, which is highly concentrated
matter so that the metric of space-time geometry is curved and one has to
apply Einstein's general theory of relativity. The space-time
geometry of a spherical neutron star described by a metric in
Schwarzschild coordinates has the form \cite{weinberg}
\begin{equation}
ds^2=-e^{\nu(r)}dt^2+[1-2M(r)/r]^{-1}dr^2+r^2[d\Theta^2+sin^2 \Theta
d\phi^2]
\end{equation}
The equations which determine the star structure and the geometry are,
in dimensionless forms \cite{weinberg,weber}
\begin{mathletters}
\begin{equation}
{d\hat P(\hat r r_0)\over d\hat r}=-\hat G
{[\hat\epsilon (\hat r r_0)+\hat P (\hat r r_0)][\hat M (\hat r r_0)
+4\pi a \hat r^3 \hat P(\hat r r_0)]\over \hat r^2[1-2\hat G
\hat M (\hat r r_0)/\hat r]},\label{tov1}
\end{equation}
\begin{equation}
\hat M (\hat r r_0)=4\pi a \int_0^{\hat r r_0} d\hat r^\prime
\hat r^{\prime^2} \hat \epsilon(\hat r^\prime r_0),\label{tov2}
\end{equation}
and the metric function, $\nu (r)$ is given by
\begin{equation}
{d\nu(\hat r r_0)\over d\hat r}=2\hat G {[\hat M (\hat r r_0)
+4\pi a \hat r^3 \hat P(\hat r r_0)]\over \hat r^2[1-2\hat G
\hat M (\hat r r_0)/\hat r]}.\label{tov3}
\end{equation}
\label{ov}
\end{mathletters}
In equations (\ref{ov}) the following substitutions have been made.
\begin{mathletters}
\begin{equation}
\hat \epsilon\equiv \epsilon/\epsilon_c,\quad
\hat P\equiv P/\epsilon_c,\quad\hat r\equiv r/r_0,\quad
\hat M\equiv M/M_\odot,
\label{tov4}
\end{equation}
where, with 
$ f_1=197.327 $ MeV fm and $r_0=3\times 10^{19}$ fm,
we have
\begin{equation}
a\equiv\epsilon_c r_0^3/M_\odot, \quad 
\hat G\equiv {G f_1 M_\odot\over r_0}
\label{tov5}
\end{equation}
\end{mathletters}
In the above, quantities with hats are dimensionless.
G in equation (\ref{tov5}) denotes the gravitational constant with
$G=6.707934\times 10^{-45}\;\mbox{MeV}^{-2}$.

In order to construct a stellar model, one has to integrate equations
(\ref{tov1}) to  (\ref{tov3}) from the star's center at $r=0$ with
a given central energy density $\epsilon_c$ as input until the
pressure $P(r)$ at the surface vanishes. As stated  in the last section,
with any reasonable
central density, we expect that at the center
we shall have only quark matter . 
Hence we shall be using here
the equation of state for quark matter through equation (\ref{pquark}) with
$\hat P(0)=P(\epsilon_c)$. We then integrate the TOV equations 
until the pressure
and density decrease to their critical values at radius 
$r=r_c$. For $r>r_c$, we shall have equation
of state for neutron matter where  pressure will change continuously
but the energy density will have a discontinuity at $r=r_c$. 
TOV equations with equation
of state for neutron matter shall be continued until the pressure vanishes.
This will complete the calculations for stellar 
model for hybrid ``neutron" star,
whose mass and radius can be calculated for different central densities.  

In Fig.4a, we plotted the mass of the star as a function of central energy
density to examine the stability of such stars. As may be seen from the
figure, $dM/d\epsilon_c$ starts becomeing negative around 1500 MeV/fm$^3$
after which it becomes unstable and may collapse into black holes
\cite{weinberg,shapiro} with the 
Chandrasekhar limit as $2.3 M_\odot$. This yields stable hybrid star of
mass $M\sim 2.3M_\odot$ with radius $R\sim 13.5$ km with a quark core
around 4.2 km as seen from Fig.4b. 
\begin{minipage}[t]{2.7in}
\begin{figure}
\psfig{figure=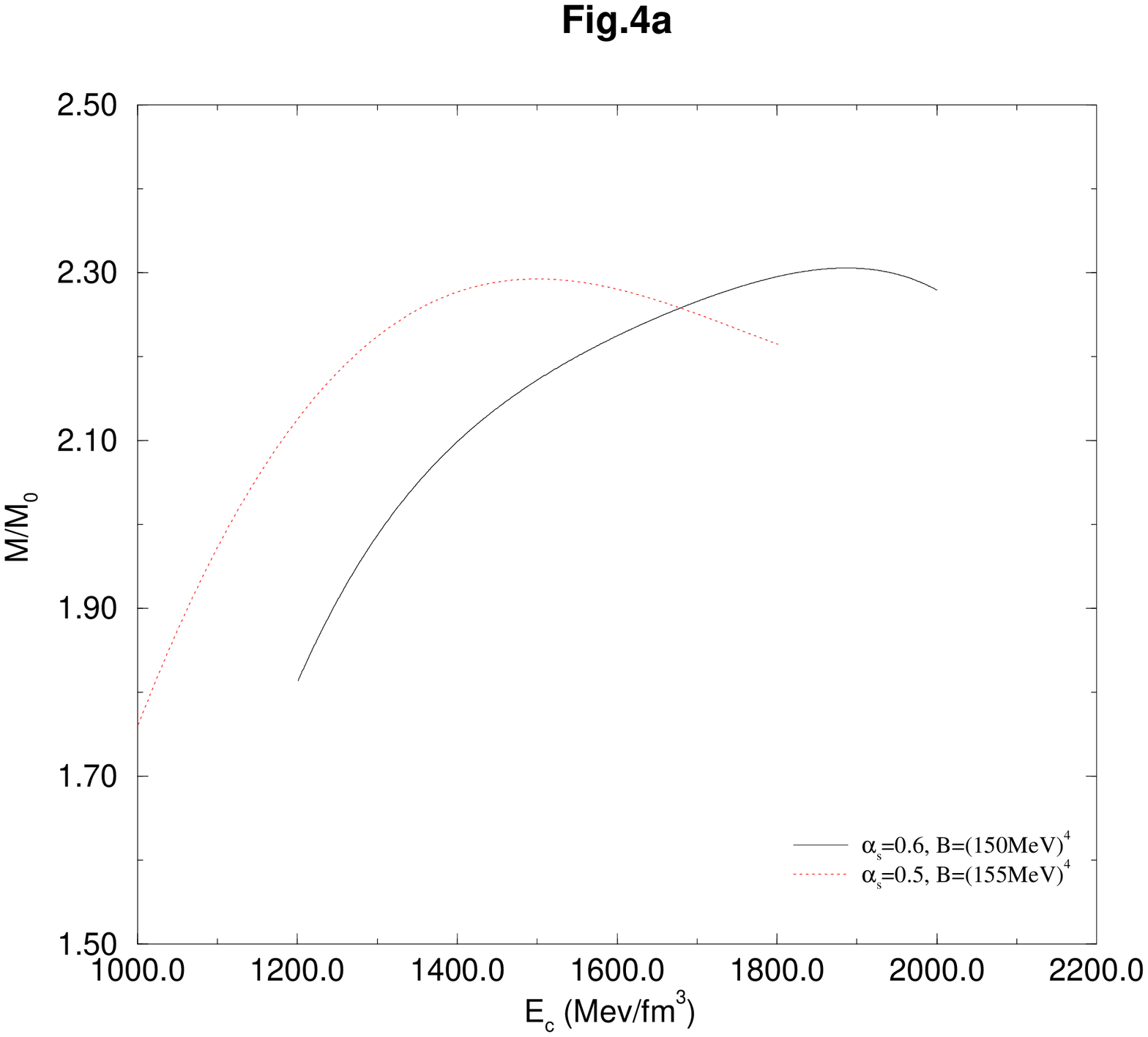,width=3in,angle=-90}
\setcounter{figure}{4}
\hspace {0.4cm}FIG. \thefigure a. The mass of the hybrid neutron star ($M/M_\odot$)
as a function of central energy density ($\epsilon_c$).
\end{figure}
\hspace{1.5in}
\end{minipage}
\begin{minipage}[t]{2.7in}
\begin{figure}
\psfig{figure=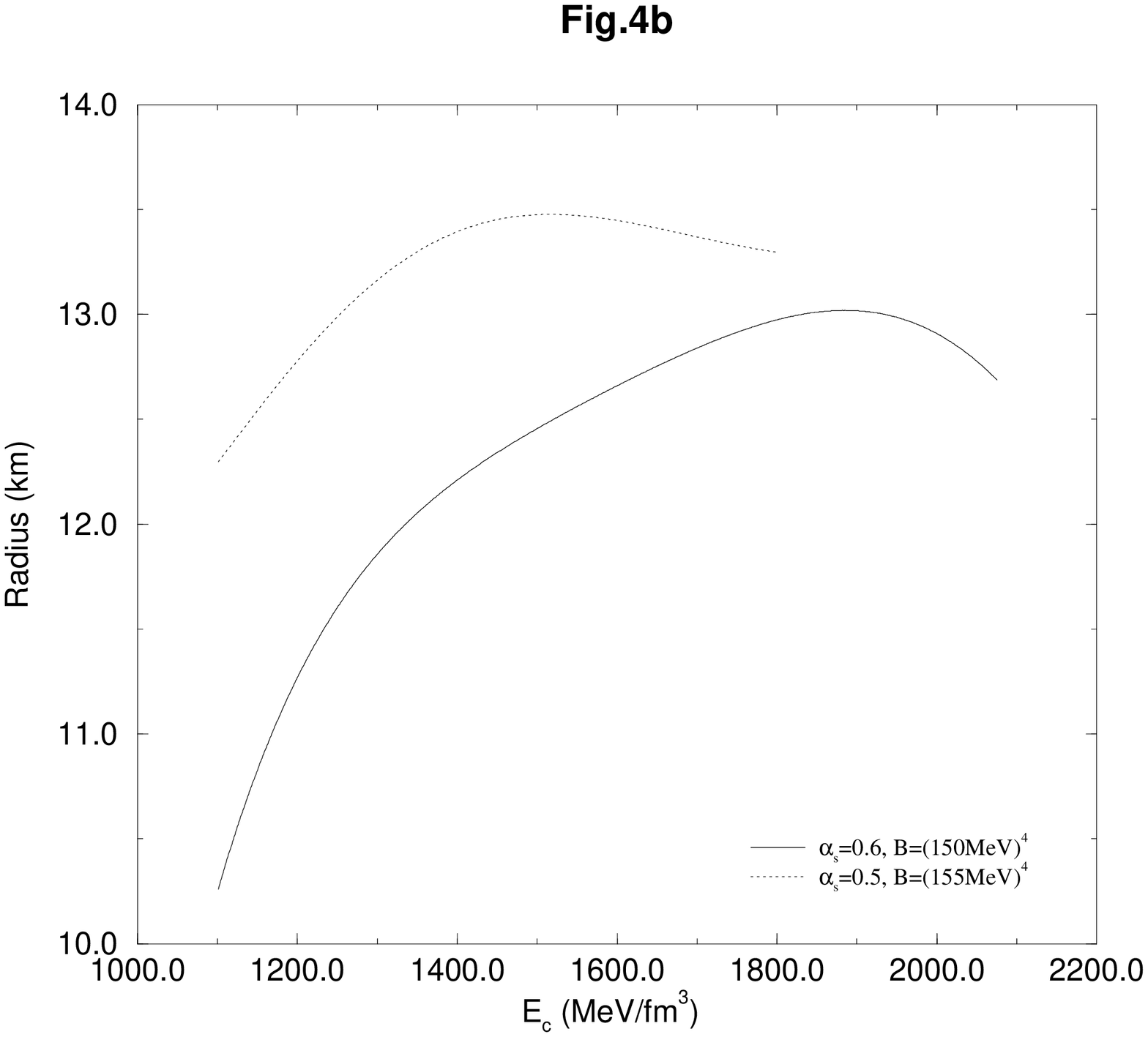,width=3in,angle=-90}
\hspace {0.4cm}FIG. \thefigure b. The mass ratio ($M/M_{\odot}$) 
versus radius (R) of the 
hybrid stars.
\end{figure}
\end{minipage}

We also calculate the surface gravitational red shift $Z_s$ of 
photons which is given by \cite{weinberg,brecher}
\begin{equation}
Z_s={1 \over\sqrt{[1-2GM/R]}} -1.\label{zs}
\end{equation}
In Fig 5 we plot $Z_s$ as a function of $M/M_\odot$. It is however 
possible that the discrete slowing down of pulsars due to the presence of two
states of matter with various mass throwing some light on the above structure.
Our graph shows there is a discontinuty of $Z_s$ around $M/M_\odot=0.4$
indicating a peculiar behaviour of redshift with Mass of the hybrid star.

Since the stars rotate about a centre, the relation between relativistic
Kepler frequency and  Newtonion Kepler frequency is given by 
\begin{mathletters}
\begin{eqnarray}
\Omega_k \simeq  0.65\Omega_c ,
\end{eqnarray}
where
\begin{eqnarray}
\Omega_c & = &\sqrt{M/R^3}\\
& =& 3.7 \times 10^3\sqrt{M/M_\odot \over{(R/km)^3}}~~~  s^{-1} .
\end{eqnarray}
\end{mathletters}

$\Omega_k$ is newtonian Kepler frequency balancing gravity with centrifugal
force . The factor 0.65 is emparical and approximate. The figures Fig.6
shows that

\begin{minipage}[t]{2.7in}
\begin{figure}
\psfig{figure=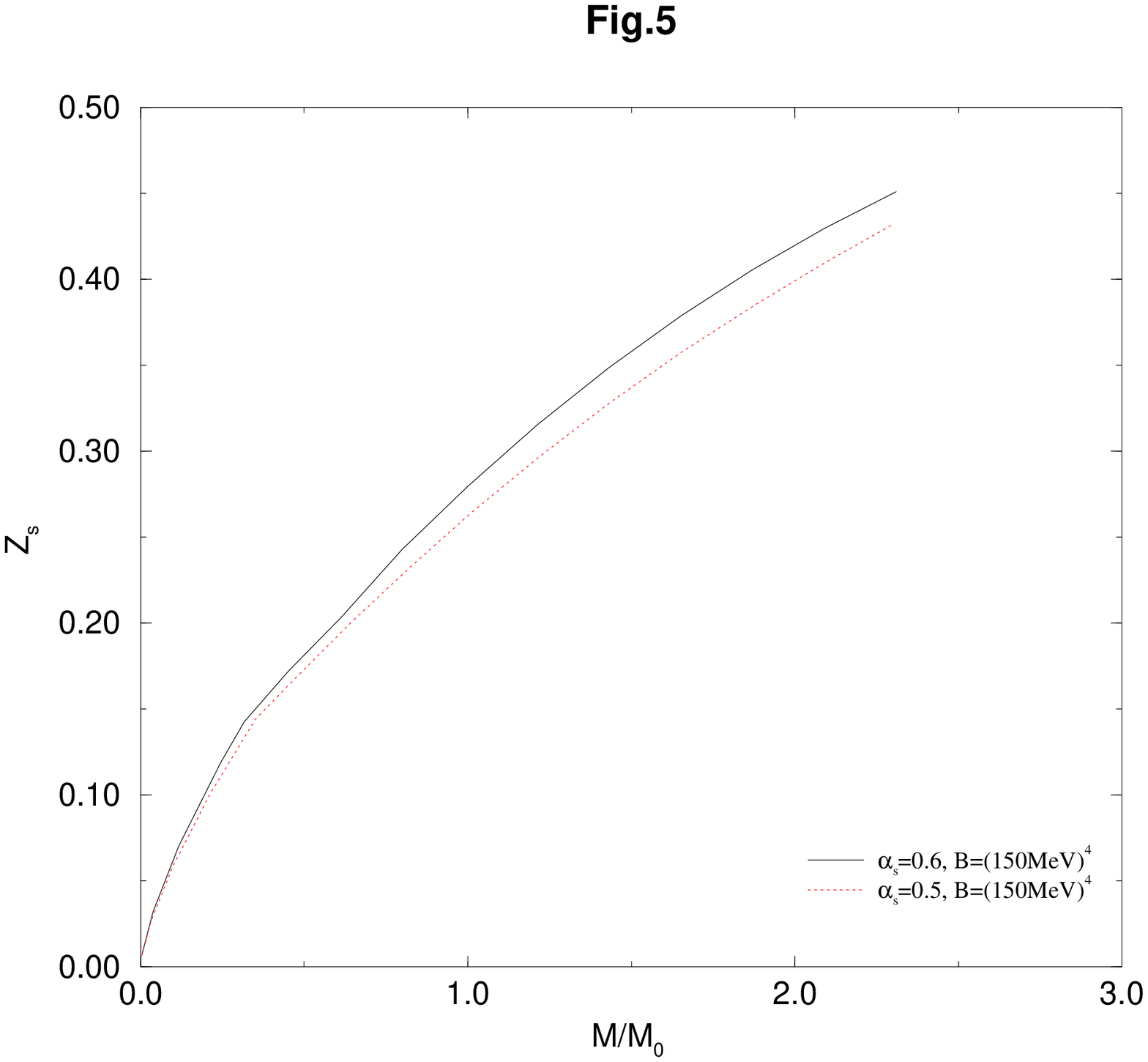,width=3in,angle=-90}
\caption{ The surface gravitational redshift ($Z_s$) as a function
of star mass ($M/M_\odot$)} 
\end{figure}
\hspace{1.5in}
\end{minipage}
\begin{minipage}[t]{2.7in}
\begin{figure}
\psfig{figure=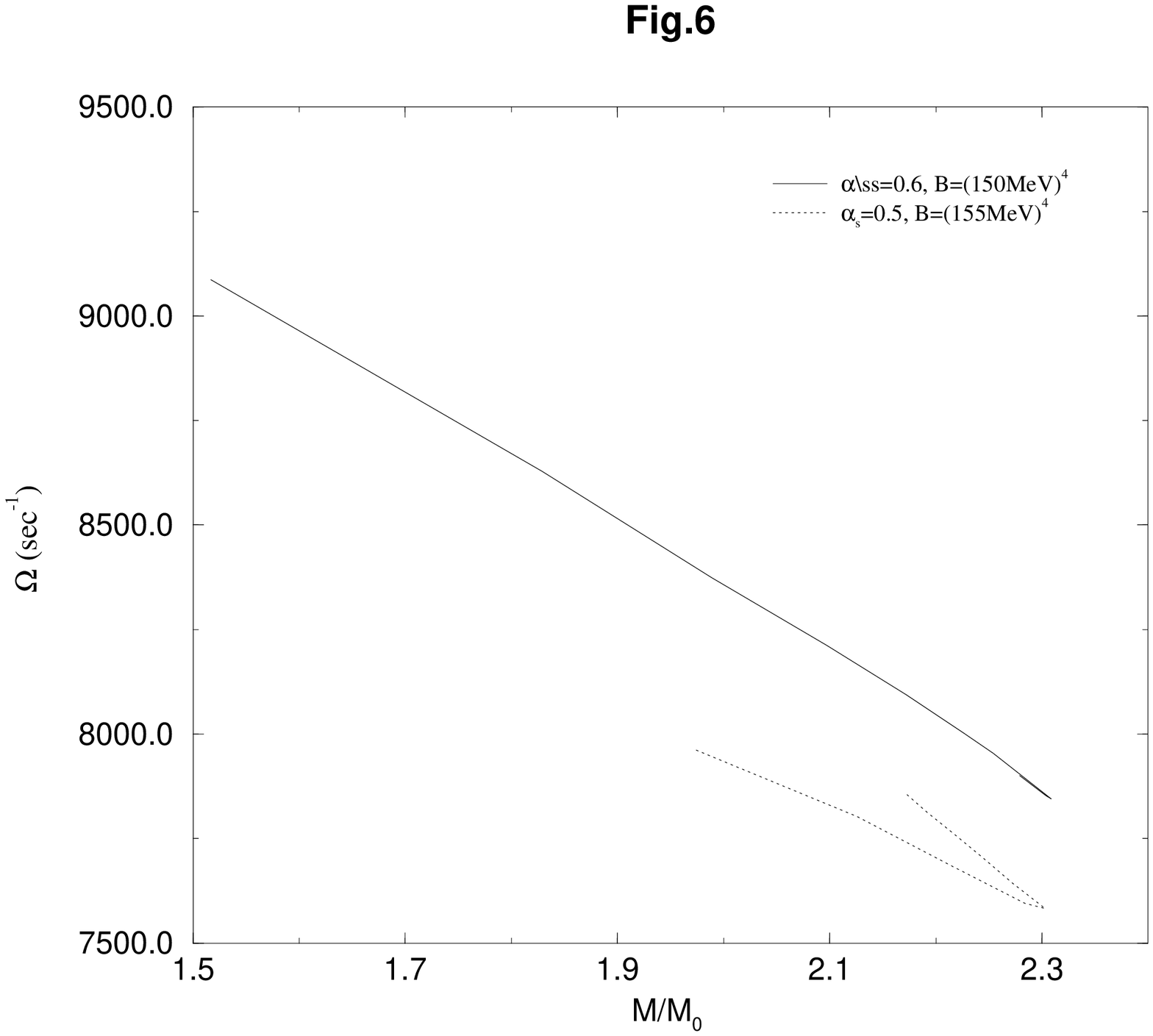,width=3in,angle=-90}
\caption{ The mass ratio ($M/M_\odot$) versus keplerian frequency 
($\Omega_c$) showing the mass ratio restricted to  2.3}
\end{figure}
\end{minipage}

the frequency is higher for lower mass and decreases for higher mass i.e.
$M/M_\odot$ is about 2.3 . It clearly shows that we can not have mass of
the hybrid star more than about 2.3 times that of sun . That also indicates
(e.g. Fig.4b) the radius of the star can not increase beyond 13.25 km .

\section{Summary and Conclusions}

We considered the equation of states taking into account the self-coupling
interactions of $\sigma$ and $\omega$- mesons. The inclusion of the 
quartic term to the $\omega$ meson field gives a soft equation of state. 
In our calculations, 
we used the TM1 parameter set, which has a capability to reproduce the
known results of finite nuclei as well as of normal nuclear matter. Here 
also the TM1 parameter set gives a phase transition for hadronic matter and
quark matter. The same parameter set predict the Chandrasekhar limit  for
Hybrid stars to be 2.3 M$_\odot$. In our calculation, we predict the inner 
quark core having a radius of about 4.2 kilometers, whereas the total 
radius of the hybrid neutron star is found to be 13.5  kilometers as compared 
to the earlier result where Chandrasekhar
limit is 1.58$M_\odot$ and radius around 10 km~\cite{mmpp}, where the nuclear 
matter equation of state was calculated through the dressing of pion pairs.
This is due to the contribution from $\sigma - \omega$ mesons. One also notes 
that the redshift has discontinuity around $M/M_{\odot}$ = 0.4, a peculiarity 
of hybrid stars. It is also observed that the Newtonian Kepler frequency of 
the hybrid stars can not increase beyond $M/M_{\odot}$ = 2.3, showing a 
decrease with increase in $M/M_{\odot}$. Pulsars are expected to be stars of 
this type but the gross properties appear to be similar to what we believe 
regarding neutron stars.

\section{Acknowledgements}

LM would like to thank DAAD Germany for 
financial help and acknowledges the hospitality at FP Physik, Universit\"at
Kaiserslautern, Germany. The help of the Computer centre at Institute of 
Physics,Bhubaneswer is duely acknowledged .

\end{document}